\documentclass[lettersize,journal]{IEEEtran}
\ifCLASSINFOpdf
\else
\fi 
\usepackage{fancyhdr}   %allows own heading definition.
\usepackage{amsmath}
\usepackage{amssymb}
\usepackage{amsthm}
\usepackage{color}
\usepackage{texdraw}
\usepackage{longtable}
\usepackage{palatino} %selects font type: palatino,times,charter,utopia,pifont,chancery,bookman,avant,helvet(ica),zapfchan(cery),courier,newcent(ury) and computer modern
\usepackage{graphicx}
\usepackage{colortbl}
\usepackage{framed}
\usepackage{cite}
\usepackage{lastpage}
\usepackage[T1]{fontenc}
\usepackage{url}
\usepackage{graphicx}
\usepackage{amsfonts}
\usepackage{listings}
\usepackage{color}
\usepackage{textcomp}

\usepackage{subfig}
\usepackage{multirow}
\usepackage{hhline}
\usepackage{array}
\usepackage{placeins}
\usepackage{makecell}  

\usepackage{glossaries}
\usepackage{tabularx}

\usepackage{algorithm}
\usepackage{algpseudocode}
\usepackage{cite}
\usepackage{threeparttable}
\usepackage{bm}
\usepackage{gensymb}
\usepackage{float}
\newacronym{APAS}{APAS}{Almost Perfect auto-correlation Sequences}
\newacronym{MPS}{MPS}{Minimum Peak Sidelobe} 
\newacronym{PMCW}{PMCW}{Phase Modulated Continuous Wave}
\newacronym{MF}{MF}{Merit Factor}
\newacronym{SNR}{SNR}{signal to noise ratio}
\newacronym{INR}{INR}{Interference to Noise Ratio}
\newacronym{SINR}{SINR}{signal to interference plus noise ratio}
\newacronym{AF}{AF}{ambiguity function}
\newacronym{MIMO}{MIMO}{Multiple Input Multiple Output}
\newacronym{SISO}{SISO}{Single Input Single Output}
\newacronym{CD}{CD}{Coordinate Descent}
\newacronym{LEO}{LEO}{Low Earth Orbit}
\newacronym{JSC}{ISAC}{integrated sensing and communications}
\newacronym{UAV}{UAV}{unmanned aerial vehicles}
\newacronym{RSMA}{RSMA}{rate-splitting multiple access}
\newacronym{DFRC}{DFRC}{dual-functional radar communications}
\newacronym{ISAC}{ISAC}{integrated sensing and communications}
\newacronym{RF}{RF}{radio frequency}
\newacronym{ISL}{ISL}{inter-satellite link}
\begin{document}
%
% paper title
% Titles are generally capitalized except for words such as a, an, and, as,
% at, but, by, for, in, nor, of, on, or, the, to and up, which are usually
% not capitalized unless they are the first or last word of the title.
% Linebreaks \\ can be used within to get better formatting as desired.
% Do not put math or special symbols in the title.
\title{Integrated Sensing and Communications Enabled Low Earth Orbit Satellite Systems}
%
%
% author names and IEEE memberships
% note positions of commas and nonbreaking spaces ( ~ ) LaTeX will not break
% a structure at a ~ so this keeps an author's name from being broken across
% two lines.
% use \thanks{} to gain access to the first footnote area
% a separate \thanks must be used for each paragraph as LaTeX2e's \thanks
% was not built to handle multiple paragraphs
%

\author{Longfei Yin,~
Ziang Liu,~
        Bhavani Shankar M. R.,~\IEEEmembership{Senior Member,~IEEE,}\\
         Mohammad Alaee-Kerahroodi,~\IEEEmembership{Member,~IEEE,}
        and~Bruno~Clerckx,~\IEEEmembership{Fellow,~IEEE}% <-this % stops a space
\thanks{L. Yin, Z. Liu and B. Clerckx are with the Department of Electrical and Electronic Engineering, Imperial College London, London SW7 2AZ, UK }% <-this % stops a space
% \thanks{B. Clerckx is with the Department of Electrical and Electronic Engineering at Imperial College London, London SW7 2AZ, UK and with Silicon Austria Labs (SAL), Graz A-8010, Austria}
\thanks{Bhavani Shankar M. R. and Mohammad Alaee-Kerahroodi are with SnT, University of Luxembourg.}% <-this % stops a space
%\thanks{Manuscript received April 19, 2005; revised August 26, 2015.}
}

\maketitle

% As a general rule, do not put math, special symbols or citations
% in the abstract or keywords.
 
\begin{abstract}
Extreme crowding of electromagnetic spectrum in recent years has led to the  challenges in designing sensing and communications systems. 
Both systems require a broad range of bandwidth, 
% to provide a designated quality-of-service 
thus resulting in competing interests in exploiting the spectrum. 
{Efficient spectrum and hardware utilization {have} led to the emergence of \gls{JSC} systems, which {have}
recently emerged as a candidate 6G technology. }
%The work would delve on different scenarios, explores the communication and sensing elements for such scenarios and brings out the \gls{JSC} paradigms along with the advantages. 
% In particular, two aspects of the \gls{JSC}: opportunistic and optimized will be highlighted in this work through the use of LEO satellites for positioning as well as the use of \gls{RSMA} technique optimized to address sensing and communication requirements.
{In particular, we provide potential techniques, namely the opportunistic ISAC and optimized ISAC.
Rate-splitting multiple access (RSMA) is highlighted as an optimized ISAC technique for LEO-ISAC
systems due to its effectiveness in simultaneously managing interference and enabling better communication-sensing trade-off performance.}
\end{abstract}

% Note that keywords are not normally used for peerreview papers.
% \begin{IEEEkeywords}
% Integrated Sensing and Communications, Low Earth Orbit, opportunistic sensing, \gls{LEO} positioning, optimized \gls{JSC}, \gls{RSMA}, \gls{MIMO} Radar, multi-user \gls{MIMO} communications
% \end{IEEEkeywords}
%
\IEEEpeerreviewmaketitle

\section{Introduction}
 \label{sec:Intro}
\IEEEPARstart{T}{he} sharing of the frequency bands between sensing and communication systems has recently received considerable attention from both industry and academia, thereby  motivating the research on systems variously known as radar-communication (Rad-Com), \gls{DFRC}, joint sensing and communications (JSC), or \gls{JSC} for 
the future beyond
5G (B5G) and 6G wireless networks. 
Techniques under the \gls{JSC} umbrella focus on designing joint systems that can simultaneously perform wireless communication and remote sensing. 
The functionalities can be combined via shared use of the spectrum, the hardware platform, and a joint signal processing framework \cite{8999605}. 
{The integrated system can either 
use the set-up of one service to enable the other in an opportunistic manner, 
or 
be optimized towards meeting the requirements of both services.} \gls{JSC} has found applicability in numerous promising applications, including autonomous vehicles, indoor positioning, etc \cite{8999605,8246850}. 
% In addition, the \gls{ISAC} enables the the co-existence of communications and radar on identical spectrum, while relying on information exchange between the two systems towards minimizing mutual interference. 
% The focus of this work will be on the {\em Opportunistic} \gls{JSC} and {\em Optimized} \gls{JSC}.

%Although remote sensing and communication technologies are developing rapidly, there is still a long way to go to realize 
Towards realizing the aim of ubiquitous and high-capacity global connectivity, satellites are envisioned to play a pivotal role to enhance the availability in unserved (e.g., deserts, oceans, forests) or
underserved areas (e.g., rural areas), enabling service reliability by providing service continuity for Internet of Things (IoT) devices or passengers on board moving platforms, and offer an infrastructure resilient to natural disasters on the ground.
% While geostationary (GEO) satellites offer all time visibility, they suffer from a large round-trip-delay causing enhanced latency not suitable for many applications. 
Compared with the Geostationary and Medium Earth Orbit (GEO/MEO) satellites which suffer from a large round-trip-delay, the 
LEO satellite systems offer a solution for lower latency, higher
data rates, and near-complete visibility.
% albeit, at the cost of larger infrastructure and complex terminals. 
Thus, LEO satellites are being considered for the next-generation wireless networks as base stations (BSs) in the sky. On the other hand, remote sensing satellites have also been considered in LEO due to their vicinity to earth.

\begin{figure*}
\vspace{-0.2cm} 
\centering
\includegraphics[width=1.4\columnwidth]{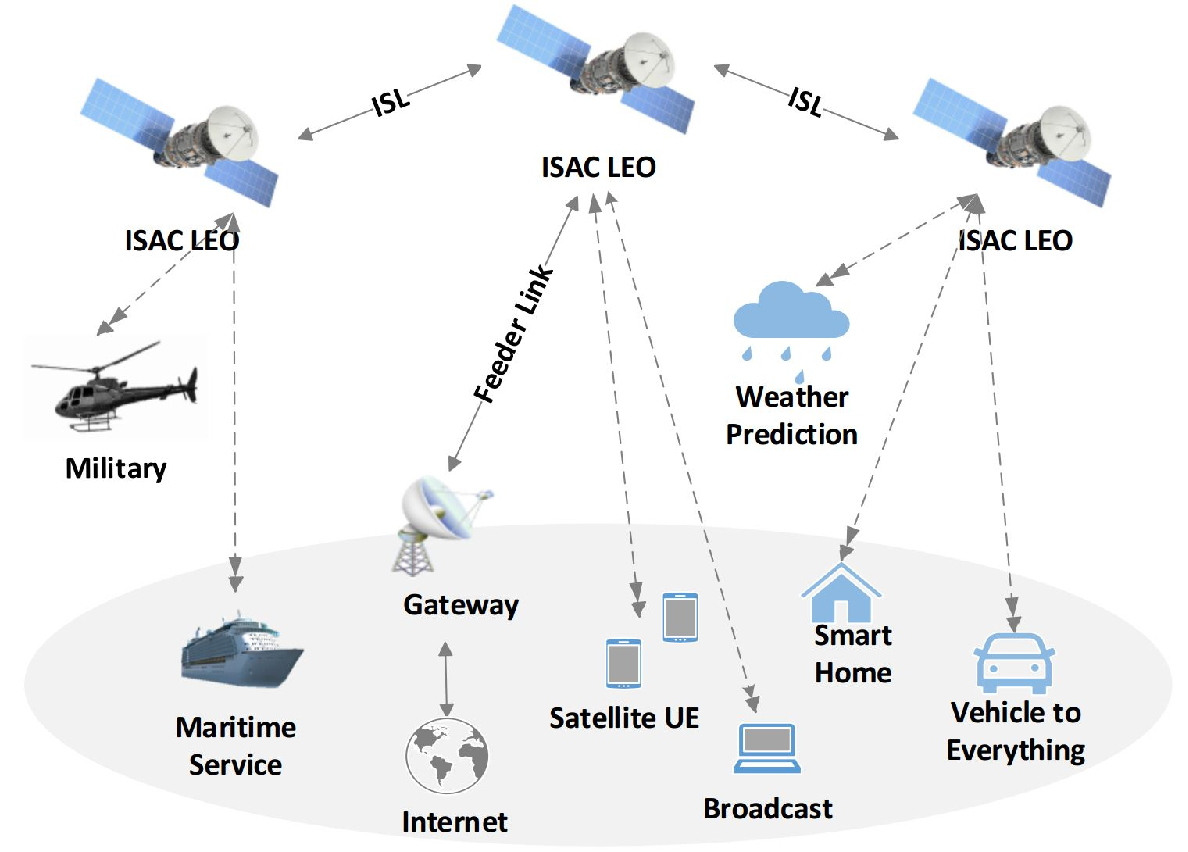}
\caption{{Model of an ISAC-enabled LEO
satellite system.}
}
\label{overviewfig}
\end{figure*}

{
The emergence of LEO satellite systems for both communications and sensing, coupled with orbit and spectral constraints, naturally motivate the ISAC in space. 
The concept of ISAC for satellite systems can be adapted to various satellite orbits according to different mission objectives. 
% The design of ISAC in each satellite type will involve practical
% considerations like communication protocols, data storage, processing capabilities, and payload design. 
However, the LEO satellite systems offer certain
advantages in ISAC compared with MEO/GEO, due to their lower latency, lower cost, and ease of integration characteristics.
Therefore, we focus on the ISAC-enabled LEO satellite systems in this article.
}
% The emergence of LEO satellite systems for both communications and sensing, coupled with orbit and spectral constraints, naturally motivate the ISAC for LEO. 
% Therefore, this article focuses on the ISAC-enabled LEO satellite systems.
% , leveraging on their lower latency, density, coverage and ease of integration characteristics. 
This scheme facilitates the integration of communications and radar sensing in LEO systems to make better use of the resources, the scarce \gls{RF} spectrum and expensive satellite infrastructure. 
{ It is worth mentioning that although there exists rich literature on LEO satellite communication systems and Rad-Com/DFRC/JSC/ISAC, there is not yet a tutorial article that consolidates the potential of applying ISAC in LEO satellite systems with various use cases as depicted in Fig. \ref{overviewfig}.
Fig. \ref{overviewfig} shows the model of an ISAC-enabled LEO satellite system which
includes two potential scenarios namely, the opportunistic ISAC
and optimized ISAC in this article.
The opportunistic ISAC uses LEO signals
as the signals of opportunities (SoO), i.e., allowing existing
communication networks to provide sensing and surveillance services.
The optimized ISAC 
focuses on optimizing the LEO
transmissions
to achieve a satisfactory
trade-off between sensing and communications.}

The rest of this article is organized as follows.
We at first present the current status of LEO satellites for both remote sensing and communications, and the motivation of ISAC in LEO. Next, we discuss the opportunistic \gls{JSC} enabled LEO satellite systems, which use LEO signals as the signals of opportunities (SoO).
% i.e., allow existing communication networks to provide sensing and surveillance services.
Two prospective scenarios are considered, namely the positioning, navigation, timing (PNT) and weather sensing.
Subsequently, we introduce the optimized \gls{JSC} enabled LEO satellite systems, where the aim is to achieve a satisfactory trade-off between sensing and communications. 
Finally, we discuss the open challenges, future directions and conclude the paper.
% focus of this article will be on the {\em Opportunistic} \gls{JSC} and {\em Optimized} \gls{JSC} enabled LEO satellite systems.
 % The waveform optimization is designed based on a rate-splitting multiple access (RSMA)-assisted multi-antenna DFRC framework.
%

% \begin{figure*}
% \vspace{-0.2cm} 
% \centering
% \includegraphics[width=1.25 \columnwidth]{Overview.eps}
% \caption{Model of a joint sensing and communications enabled LEO
% satellite system.
% }
% \label{overviewfig}
% \end{figure*} 
% %
% \vspace{-0.2cm}
% \section{LEO Satellite Systems: Current Status and Motivation for ISAC}
% \label{sec:LEO}

\vspace{-0.2cm}
\section{LEO Satellite Systems: Current Status and Motivation for ISAC}
\label{sec:LEO}
Recent years have witnessed 
increased attention in the industry towards LEO satellites.
A large number of satellite operators have announced plans to develop dense small satellite networks by launching thousands of low-cost LEO satellites to satisfy the growing demands for high capacity, ultra-reliable and low latency connectivity, seamless coverage, IoT, earth observation, and wireless sensor networks (WSNs), etc.
%LEO satellite systems are capaoffer  a wide range of services,
%including Earth observation and synthetic aperture radar (EOSAR) applications (e.g., ICEYE, HawkEye, and Satellogic), weather forecasting (e.g.,  NOAA 15, 18-20, MetOp, Fengyu-3), IoT applications (e.g., Hiber, Myriota, and Astrocast), and broadband connectivity (e.g., Iridium, Globalstar, OneWeb, and Starlink) \cite{prol2022position}, etc.
This section provides an overview of the current status of LEO satellite systems in the topic of interest of the paper $-$  remote sensing and communications, and motivates  \gls{JSC} in LEO systems.
\vspace*{-0.1in}
\subsection{LEO Remote Sensing}
Remote sensing involves the contactless acquisition of information about an object or phenomenon by measuring its reflected and emitted radiation from a distance. In the past decade, LEO satellites have been used for remote sensing in numerous fields, including geography, ecology, agriculture, meteorology, oceanography, military, commercial, extreme weather forecasting and disaster monitoring, etc. The lower operational altitude of LEO satellites offers improvements in radiometric performance (signal-to-noise ratio), geospatial position accuracy, communication link budgets, and lowers the risk of collision with space debris compared to GEO and MEO orbits \cite{crisp2020benefits}.
%The decrease in antenna areas and power requirements, and the development and launch of smaller spacecrafts can lead to a lower-cost system.
Besides, the LEO constellation design is able to provide larger area coverage and thereby improving the temporal resolution of imagery or data \cite{su2019broadband}.

The emerging LEO remote sensing typically focuses on Earth monitoring and weather forecasting applications. The ICEYE,  through its X-band Synthetic Aperture Radar (SAR) satellites, aims to provide high-resolution view of the Earth's surface with the ability to observe a location at different times of the day. 
%and  is revolutionising the frequency of global imaging with which we can observe and react to changes in near real-time of any location. 
Due to the SAR capabilities, ICEYE is  able to provide high-resolution radar images under low visibility conditions regardless of weather conditions and time of day.

In terms of LEO weather forecasting,  sun-synchronous polar orbiting meteorological satellites are designed, which pass over the poles and view every location twice a day. The MetOp series, designed on behalf of Eumetsat in 2006, is the first  Europe's polar-orbiting satellite for observational meteorology.  Polar orbiting weather satellites offer a much better resolution than the geostationary counterparts due to their vicinity to Earth.
\vspace*{-0.1in}
\subsection{LEO Communications}
LEO satellite communication is very attractive because of its relatively lower latency, path loss and production and launching cost, in comparison to GEO and MEO satellites. With the rapid development of relevant technologies, the weight and volume of the satellite user terminals (UTs) are similar to those of personal mobile devices, and could potentially be integrated with the latter; this enhances the commercialization. The space-based LEO network architecture has inter-satellite links (ISLs) to reduce the number of on-ground gateways. By employing onboard processing (OBP) regenerative payloads, the ISL links enable communication and cooperation among satellites, data transmission between UTs, data routing, throughput maximization, latency minimization and seamless coverage, etc \cite{hassan2020dense}. Moreover, LEO constellations comprising of multiple satellites can achieve global coverage for broadband connectivity and an easy-to-access infrastructure. Various strategies such as smart steerable satellite antenna design, multiple access, routing and resource algorithms further boost the  LEO communications capability.

Supported by the developments in aerospace and electronic information technology, and aided by the joint promotion of the government and commercial companies, LEO communications has moved from the conceptual to the operational stage. OneWeb, an emerging LEO communication constellation, aims to provide affordable network connectivity for remote areas or areas with poor Internet infrastructure. 
{ It would contain 822 satellites (648 active/ 234 redundant) evenly placed on different polar orbital planes at around 1200km from the ground, using Ku and Ka band \cite{al2022survey}}.
%The satellites move at a high speed, and different satellites appear in the sky alternately to ensure the signal coverage of a certain area.
% The goal  is to provide global broadband internet services to people everywhere by the end of 2023.
Another constellation under operation is the Starlink proposed by SpaceX, which started launching satellites in 2019 with the aim of providing users with ultra-high speed communication with a minimum data rate of 1 Gbps and a maximum rate of 23 Gbps. {In total, nearly 12,000 satellites are planned to be deployed, with a possible later extension to 42,000 \cite{al2022survey}}. 
%All satellites are able to autonomously avoid collisions based on uplinked tracking data, and are equipped with thrusters which allow them to de-orbit at the end of their life. 
The entire system offers a high degree of flexibility, wherein the signals can be dynamically concentrated to specific regions, so as to provide high-quality network services.
%
%{ \section{Joint Sensing and Communications  Enabled LEO Satellite Systems}
\vspace{-0.2cm}
\subsection{ISAC-enabled LEO Satellite Systems}
%To fulfill the increasing demands on high-quality wireless connectivity as well as accurate and robust sensing capability, joint sensing and communications has been  envisioned as a key technique for future 6G wireless networks to improve the utilization efficiency of the  spectrum resources, signalling and hardware platforms. The existing joint sensing and communications works mainly focus on terrestrial networks with novel use cases including autonomous vehicles, smart cities, smart homes, human computer interaction, smart manufacturing and industrial IoT, etc \cite{liu2022integrated}. According to the aforementioned discussions about LEO remote sensing and LEO communications industrial activities, joint sensing and communications enabled LEO satellite system is envisioned to reduce the number of required satellites, save orbit and frequency resources, and reduce the launch operation and maintenance costs. In the following Section III and Section IV,  two aspects of joint sensing and communications enabled LEO satellite systems will be discussed, namely the Opportunistic JSC and Optimized JSC. 
%
To fulfill the increasing demands on high-quality wireless connectivity as well as accurate and robust sensing capability,
ISAC has been envisioned as a key technique for future 6G wireless networks to improve the utilization efficiency of the
spectrum resources, signalling and hardware platforms. The existing works on ISAC  mainly focus on terrestrial networks with novel use cases including autonomous vehicles, smart cities, smart homes, human-computer interaction, smart manufacturing and industrial IoT, etc \cite{8999605}. According to the aforementioned discussions about LEO remote sensing and LEO communications, ISAC-enabled LEO satellite system is envisioned to reduce the number of required satellites, save orbit and frequency resources, and reduce the launch operation and maintenance costs. In the following Section III and Section IV,  two aspects of ISAC-enabled LEO satellite systems will be discussed, namely the Opportunistic ISAC and Optimized ISAC. 
%{\color{red} Remark: two subsections are as follows: LEO-PNT and LEO-ISAC. in each subsection, there will be some subsubsections.
%}
\section{Opportunistic ISAC}
In this section, we focus on the Opportunistic ISAC, where existing LEO wireless signals are used as the signals of opportunities
(SoO) to provide sensing and surveillance services
such as PNT and weather sensing.
% , while no specifically designed PNT or weather sensing signals are transmitted.
% LEO wireless signals are used as the signals of opportunities (SoO) for remote sensing, e.g., positioning, navigation, timing (PNT) and weather sensing, while no specifically designed PNT or weather sensing signals are transmitted.
%
\subsection{LEO-PNT}
Resilient and accurate PNT is of paramount importance in safety critical systems such
as aviation and transportation. The recent unsolved case of the Malaysian Airlines MH-350 is a classical example highlighting this need. As systems evolve towards becoming fully autonomous, the requirements on their PNT become ever more stringent. With no human in-the-loop, an inaccurate PNT solution or the PNT system failure could have catastrophic consequences. Until now, PNT services are mainly offered by global navigation satellite system (GNSS), which relies on MEO satellite systems, such as the Galileo, GPS, Glonass, and Beidou. 
However,  GNSS  are mostly located at altitudes of 19100 km and beyond, resulting in high attenuation and, hence, the GNSS cannot function satisfactorily indoors and in some challenging outdoor scenarios. Additionally, the updates of parameters 
%navigation and positioning speed 
is slow when the number of satellites is insufficient.  In such contexts, current vehicular navigation systems couple GNSS receivers with an inertial navigation system (INS) to benefit from their complementary properties: the short-term accuracy and high data rates of an INS and the long-term stability of a GNSS PNT solution to provide periodic corrections. However, in the inevitable event that GNSS signals become unreliable (e.g., in deep urban canyons or near dense foliage), unusable (e.g., due to unintentional interference or intentional jamming), or untrustworthy (e.g., due to malicious spoofing attacks or system malfunctions), the navigation system relies on unaided inertial measurement unit (IMU) data.
As a consequence, the errors accumulate and eventually diverge the vehicle’s efficient and safe operation.

% Signals of opportunity  could be used in GNSS challenged environments  Classical signals include AM/ FM radio, cellular, digital television, and low Earth orbit (LEO) satellites. 
Recently, there has been an increased research effort towards exploiting the LEO potential in the context of PNT. The literature considers LEO-PNT  attractive in some GNSS denied environments for the following reasons:
% \cite{UCI_PNT}:
%
\begin{itemize}
    \item LEO satellites are around 20 times closer to Earth compared to GNSS, thereby resulting in lower latency and better signal propagation conditions.
    \item LEO satellites orbit the Earth much faster than GNSS satellites, making their Doppler measurements attractive to exploit.
    \item The recent announcements by OneWeb, Boeing,  Starlink, Samsung, Kepler, Telesat, and LeoSat to provide broadband internet via  LEO satellites will collectively bring thousands of such satellites into operation, making their signals abundant and diverse in frequency and direction.
   \item Multiple satellites in a LEO constellation offer global coverage (including polar regions where GNSS signals are weak) and three-dimensional (3-D) location information along with precise timing. % Further, the polar orbits ensure global coverage, including at the poles (where the GNSS signals are weak).
  %  \item 
   % At their lower orbit altitude, LEO satellites are well protected from space disturbances (e.g., solar storms), and they are far less susceptible to the deleterious effects of internal charging and surface charging that can cause permanent damage to the electronic components of space vehicles operating in MEO or GEO orbits.
    %LEO satellites are better shielded from these phenomena because the orbits are below the Earth’s magnetosphere.
   %They have low construction cost and fast startup speed, which can be used as a supplementary backup for the medium and high orbit PNT systems.
\end{itemize}
Two measurement methodologies can be used to determine the position of a receiver. 
\paragraph{Doppler Based} Here, the Doppler perceived at the receiver is related to the position of the receiver as well as the position and velocity of the satellite. Using the publicly available data and motion models, the satellite position and velocity can be determined within a certain accuracy. Further, the receiver determines the Doppler relying on the communication protocol (pilots) and classical frequency estimation techniques. Subsequently, a tracking algorithm like the Extended Kalman Filter (EKF) can be used to estimate the position of the receiver using multiple measurements from a satellite.
\paragraph{Carrier Based} In this case, the phase of the received carrier is related to the position of the receiver analytically. However, carrier phase measurements are ambiguous due to the wrapping and the technique needs a reference node to overcome the ambiguity.

Our recent work has explored the LEO-PNT with an enhanced focus on satellite system aspects and has indicated that the Doppler based method is very robust, and the performance improves with the number of satellites \cite{UL_PNT}. This opens up the possibility to leverage on the future 6G deployment, which has protocol support from both NTN and positioning, and offers a low-cost method to supplement/ complement existing PNT mechanisms.
\subsection{LEO-Weather Sensing}
LEO satellite based cloud and precipitation radars  have established themselves  as an excellent tool for providing 3-D structure of rainfall, particularly the vertical distribution, and obtaining quantitative rainfall measurements over land and ocean.  In contrast to ground-based weather radars, which typically operate at S, C, and X-band frequencies, Ku band is currently the lowest radar frequency used in space, with Ka and W as alternatives\footnote{The W band frequency is the preferred cloud sensing frequency due to its sensitivity to detect the radiatively and hydrologically important clouds.  However, a multifrequency approach observing the same precipitating system with different operating frequencies is typically used.}
%in LEO-based weather radar systems.} 
\cite{https://doi.org/10.1029/2019RG000686}.   
They are  further used  for mapping of most types of meteorological events in all locations of the globe,
where installation of ground based radars is either impossible or infeasible.
% \cite{8378741}.

Traditionally LEO-based weather radars, e.g., TRMM,  GPM, CloudSat, and EarthCARE, have been developed as an extension of ground based and airborne weather radars, by utilizing high transmit power and large antenna gain.
% The typical scanning scheme of LEO-weather radar is based on phased-array systems, and mechanical scanning is not practical due to fast platform motion.
Phased-array technology is currently being used to miniaturize these fascinating radars so they may be used in small 6U spacecraft like CubeSats, increasing the temporal sampling frequency of the entire planet. The challenge with this new development is to simultaneously miniaturize, cut costs, and retain the fundamental performance standards for this kind of radar, which encourages the employment of unique architectures outfitted with optimal waveforms and pulse compression algorithms \cite{https://doi.org/10.1029/2019RG000686}.
% Indeed, the radar equation shows that for distributed targets such as precipitation, the received power is proportional to the length of the transmit pulse \cite{sadowy2003development}.
% As a result, pulse compression, which improves sensitivity while maintaining the resolution of conventional short pulse radars, can improve performance, i.e., lower antenna gain or transmit power is required for comparable sensitivity.
However, pulse compression has a drawback in applications involving weather sensing. The return from rain slightly above the surface may be obscured by range sidelobes in the matched filter output from nearby cells where there is a strong surface, such as ground. In this case, a joint design of the transmit waveform and the extended mismatch filter can be used, resulting in pulse compression with almost no sidelobes and a high signal-to-interference-plus-noise ratio (SINR) which guarantees a low estimation error on meteorological reflectivity \cite{9884519}. 
Further research and weather monitoring are required in areas without access to specially built weather sensors due to ongoing urbanization and climate change.
In order to be employed as an opportunistic sensor, as was previously accomplished successfully via satellite terminals and commercial microwave links, LEO-based weather radars' next step must be their integration with communication.

% {\bf \color{red} @Mohammad: Something on the trend towards opportunisitic?}
% Higher radar frequencies (e.g., G band and submillimeter wave) have not been used yet but offer an attractive
% solution for dual‐wavelength measurements at small particle sizes and for water vapor profiling. 
% \section{Optimized ISAC: LEO-ISAC}
% %
% \begin{figure*}
% \vspace{-0.5cm} 
% \centering
% \includegraphics[width=1.76 \columnwidth]{LEO-RSMA2.eps}
% \caption{RSMA-assisted LEO-ISAC.
% }
% \label{rsma}
% \end{figure*} 

\section{Optimized ISAC: LEO-ISAC}
\begin{figure*}
\vspace{-0.5cm} 
\centering
\includegraphics[width=2.06\columnwidth]{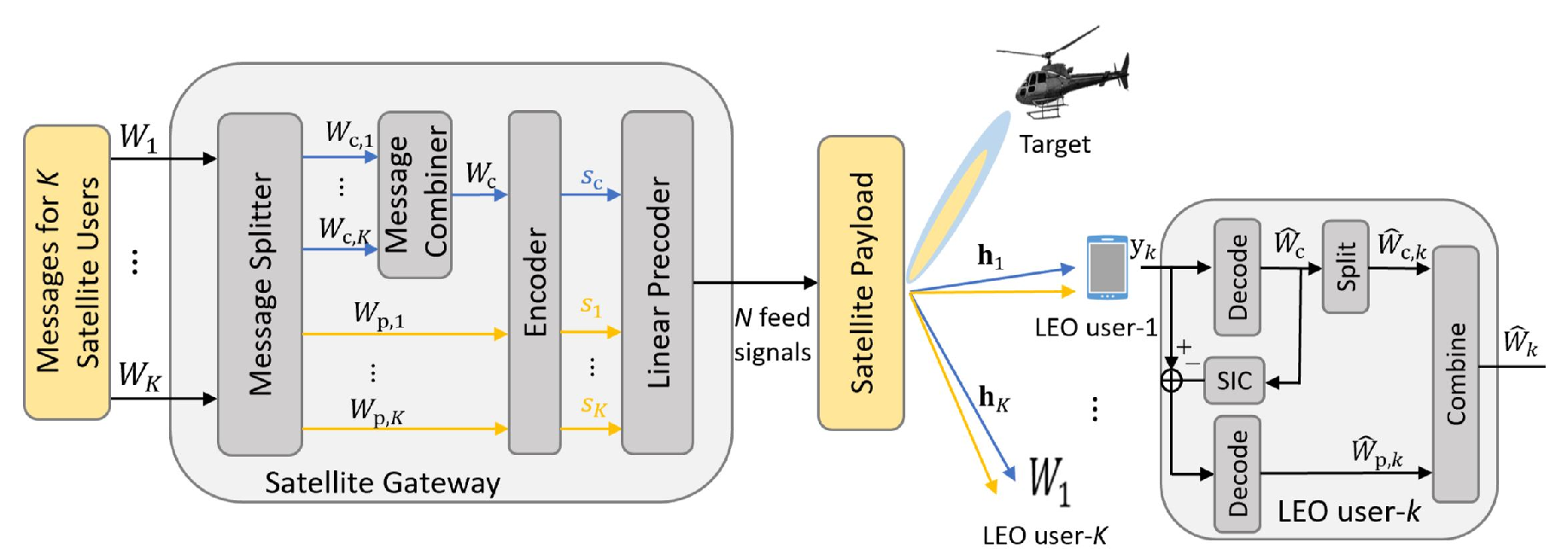}
\caption{{RSMA-assisted LEO-ISAC.}
}
\label{rsma}
\end{figure*} 
%
%In addition to the Opportunistic JSC which investigates the potential usage of existing LEO wireless signals to achieve sensing functionalities,  in this section, we focus on the Optimized JSC and present a feasible study on the LEO-ISAC scenario.  
Complementary to the previous section on the opportunistic use, this section focuses on optimizing the LEO transmissions to optimize sensing and communications in a feasible LEO-ISAC scenario.
%and presents a feasible study on the LEO-ISAC scenario.  
%The transmitted LEO signals are optimized to achieve satisfactory performance in both sensing and communications.

As a key feature of the 6G radio access network (RAN), ISAC which merges sensing and wireless communications
into a single system, has drawn extensive research attention. Both functionalities are combined via shared use of the spectrum, the hardware platform,
and a joint signal processing framework.
% Herein, we use ISAC to refer to the joint waveform optimization to achieve satisfactory performance in both sensing and communications, rather than relying on existing waveforms.
The goal of ISAC is to unify the two functionalities and pursue direct trade-offs between them.
% as well as mutual performance gains.
Most existing ISAC works
focus on terrestrial networks \cite{8999605}.
However, terrestrial ISAC systems cannot provide global services and the data reception and processing resources are very limited.
With the continuous development of LEO satellite design and manufacturing technology, and the enhancement of on-board processing capacity, some existing works have proposed to operate ISAC through LEO satellite systems. They focus on dual-functional joint waveform design/optimization, and show great potential in providing
wide coverage for both wireless communications and sensing.
{Next, we discuss some important technologies that can support LEO-ISAC and some promising application scenarios.}
\subsection{Joint Beamforming for LEO-ISAC}
% \hfill\par
Beamforming has been widely studied to enhance the performance of multi-antenna systems to overcome co-channel interference (CCI) introduced by frequency reuse when the channel state information (CSI) is available at the transmitter. 
{The rapid development of beamforming techniques in terrestrial systems has spurred its implementation 
in satellite systems to
mitigate interference and
enhance system spectral efficiency. 
The transmitted messages are spatially separated by beams that ideally steer a maximum amount of power
into a specific direction while minimizing interference amongst different beams.
}
%Due to the large propagation distances from the satellite to Earth, these smart steerable satellite antennas require sufficient antenna gains, which can counteract the free space power loss. 
Beamforming technology can be leveraged  by satellites not only for regular satellite-Earth communications but also for sensing capability. The key differences between LEO-ISAC and  the existing works on terrestrial-ISAC lie in the use of the multiple spot beam coverage scheme and the significantly different propagation properties, i.e., the high propagation delay and large Doppler shifts due to long distances between the LEO satellites and the UTs/targets as well as their mobility.

In satellite communications, LEO satellites can serve a massive number of satellite user equipments by employing multi-beam techniques, and the adoption of full frequency reuse (FFR)
scheme
is the leading paradigm to increase the overall system throughput. 
Thus, inter-beam interference becomes a major obstacle.
The LEO-ISAC  joint beamforming can be optimized to achieve a good trade-off and satisfactory performance in both sensing and communication. The trade-offs in ISAC include the information-theory limits, PHY performance and cross-layer metrics \cite{8999605}.
In terms of PHY sensing metrics, the detection probability, false-alarm probability, mean squared error (MSE), Cramer-Rao Bound (CRB), beampattern approximation and the radar mutual information (RMI), etc, are of particular interest, while the PHY communication metrics include spectral efficiency (SE), energy efficiency (EE), outage probability, bit error rate (BER), symbol error rate (SER), etc.
In \cite{9852292}, the application of ISAC was investigated in MIMO
LEO satellite systems, and the beam squint-aware
hybrid joint beamforming was optimized  to operate communications and target sensing simultaneously.
A performance trade-off was enabled between the communication EE and the sensing beampattern. 
\subsection{RSMA Technique for LEO-ISAC}
% \hfill\par

In a LEO-ISAC system, to better manage the inter-beam interference among communication
users, as well as providing satisfactory performance in sensing, advanced multiple
access techniques become indispensable.
Rate-splitting multiple access (RSMA) has been recognized as a generalized
and powerful non-orthogonal transmission framework, and robust interference management strategy for multi-antenna multi-user networks that encompasses
the conventional beamforming technique \cite{mao2022rate}, i.e., space division multiple access (SDMA) as a sub-scheme.
RSMA relies on linearly precoded rate-splitting (RS)
at the transmitter and successive interference cancellation (SIC) at the receivers.
{RSMA has recently been proven to flexibly manage different
types of interference in satellite communication systems e.g., intra-satellite system
interference, inter-satellite interference, and interference from terrestrial systems\cite{9844445}.}
Moreover, RSMA-assisted ISAC satellite
system has also been investigated in \cite{9771644}, where 
RSMA was demonstrated to be
very promising to enable a
better communication-sensing trade-off 
%and achieve better target estimation 
performance than the conventional SDMA-assisted joint beamaforming strategy.
{
As illustrated in Fig. \ref{rsma}, at the transmit side, each message is split into a common part and a private part, which are then encoded and linearly precoded
at the satellite gateway.
At the receive side, each user first decodes the common stream by treating all private streams as noise. After removing the
decoded common stream, each user
decodes the intended private stream by treating the other private streams as noise.
This enables the flexibility of RSMA by
partially decoding interference and partially treating it as noise, and thereby RSMA can reduce
to SDMA by tuning the powers and contents of the common and private streams.
The introduced additional common stream can not only better manage interference between communication users, but also act as a radar sequence to assist the sensing functionality.
}

\begin{figure}
\vspace{-0.2cm}
\centering
\includegraphics[width=1 \columnwidth]{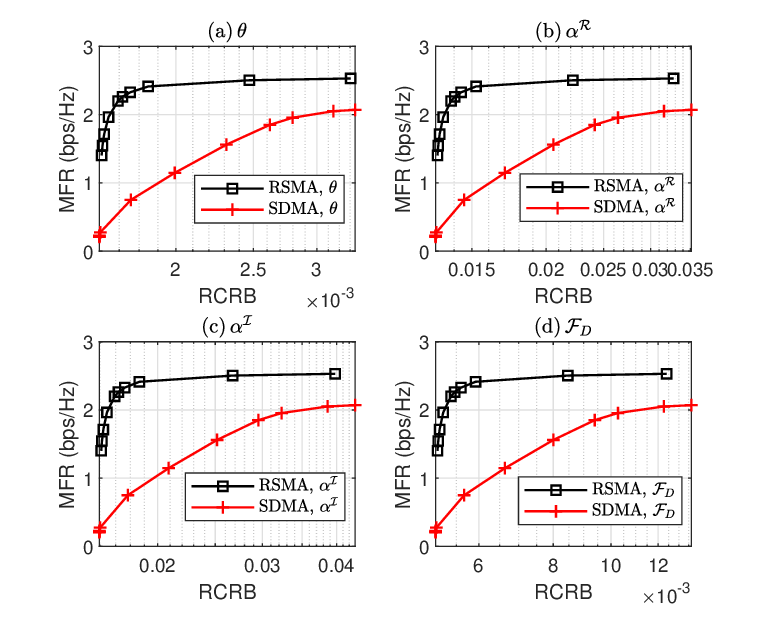}
\caption{MFR and RCRB trade-off in a LEO-ISAC system, \\(a) $\theta$, (b) $\alpha^{\mathfrak{R}}$, (c) $\alpha^{\mathfrak{I}}$, (d) $\mathcal{F}_{D}$.
$N_{t} = 8, K = 16.$
}
\label{LEOSAT}
\end{figure}

Fig. \ref{LEOSAT} depicts the optimized  trade-off performance between the minimum
fairness rate (MFR) amongst satellite users and the root Cram\'{e}r-Rao bound (RCRB) of target parameter estimation in a LEO-ISAC system with $N_{t} = 8$ satellite antenna feeds serving $K = 16$ single-antenna users. We assume $\rho = 2$ the communication users are uniformly distributed
in each satellite beam.
The target parameters $\theta$, $\alpha^{\mathfrak{R}}$, $\alpha^{\mathfrak{I}}$, $\mathcal{F}_{D}$ respectively denote
the direction of arrival (DoA)/direction of departure (DoD), the real/imaginary part of the complex
reflection coefficient and the Doppler frequency.
The performance gain achieved by RSMA-assisted LEO-ISAC joint beamforming can be clearly observed compared with the conventional SDMA-assisted joint beamforming.
% The readers can refer to \cite{9832622} for further details.
{
The benefit of using RSMA in ISAC is due to the dual purposes of the common stream as illustrated in Fig. 2. 
At the leftmost points of Fig. 3 which correspond to prioritizing the sensing functionality, both RSMA and SDMA can achieve very good sensing performance because the precoders are optimized to radiate the highest power towards the target angle. 
However, SDMA-assisted
ISAC can no longer exploit spatial degrees of freedom (DoF) provided by multiple antennas and thereby experiences lower MFR, compared with the
RSMA-assisted ISAC which has an additional DoF introduced by the common stream.
The readers can refer to \cite{9832622}, \cite{chen2023rate} for further details about the effectiveness of RSMA for multi-antenna and 
multi-target ISAC systems.
% The trade-off gain provided by RSMA grows
% with the increase of angle difference between multiple targets.
}

{
\textit{Remark:}
In RSMA-assisted ISAC systems, the spectrum sharing between
communication and radar sensing raises security concerns, especially
for military applications. 
Higher layer encryption is required.
From a
physical layer security/secrecy perspective,
RSMA and the trade-off between sensing and secrecy rate maximization are also worth being investigated.
Moreover, low complexity techniques in RSMA-assisted ISAC are challenging to be designed since the precoders and power allocation to the common stream could depend on different network scenarios and metrics.
}

{
\subsection{LEO-ISAC Application Scenarios}

Some promising application scenarios of the proposed LEO-ISAC system are given as follows:
\begin{itemize}
    \item Maritime Surveillance: LEO satellites with ISAC capabilities can effectively detect and track maritime vessels. The system can be used to monitor shipping lanes, identify illegal  activities, track vessels, and provide real-time updates on maritime traffic. It allows for continuous monitoring and real-time communication with authorities for timely response.
    \item 
    Border Management:  LEO-ISAC systems can be used for border surveillance and security. 
    LEO satellites equipped with sensing capability can detect and track movements along borders and monitor suspicious activities. The communication capability
    % enables the transmission of real-time information, 
    facilitates real-time response and effective border management.
    % \item 
    % Wildlife Conservation: LEO satellites with integrated sensing and communication systems play a crucial role in wildlife conservation efforts. They can detect and track endangered species, monitor their habitats, and identify illegal poaching activities. The real-time transmission of data allows for immediate response and intervention, enhancing protection measures and helping to combat wildlife crimes.
    % \item 
    % Disaster Monitoring and Early Warning Systems: The integrated system can contribute to the detection and tracking of natural disasters such as hurricanes, wildfires, and floods. LEO satellites equipped with sensing capabilities can monitor atmospheric conditions, detect fire hotspots, and track the movement of weather systems. By transmitting this information in real-time, early warning systems can be established, allowing communities to evacuate and emergency services to prepare accordingly.
    \item 
    Air Traffic Management: LEO-ISAC systems can assist air traffic management. 
    % It can detect and track aircraft, monitor airspace congestion, and provide real-time information on flight paths and weather conditions. Therefore, 
    It can be used to optimize air traffic flow, improve safety, and enable efficient communication between pilots and air traffic control.
    \item 
    Space Situational Awareness: LEO-ISAC systems can detect and track objects in space, such as satellites, space debris, and potential threats. The sensing capability can help to perform precise tracking and collision avoidance. The communication capability can facilitate information sharing with space agencies and satellite operators for more effective space situational awareness.
\end{itemize}
}

\vspace{-0.2cm} 
\section{Challenges and Future Directions}
Note that the research in LEO-ISAC system is in its infancy, thus there are numerous issues to address. We conclude the unsolved open challenges of LEO-ISAC systems, and provide potential solutions to these issues.

\textbf{Doppler Effect Mitigation:}
The high-speed mobility feature of the LEO-ISAC system results in a severe Doppler effect that degrades the system's link reliability. One possible solution to this issue is compensating for the frequency-selective and time-varying channel effects by using machine learning techniques to accurately estimate the channel. Another promising solution is to adopt Doppler-resilient transmission techniques, e.g., orthogonal time-frequency space (OTFS) modulation. In OTFS modulation, the doubly-dispersive channel in the time-frequency (TF) domain is transformed into the delay-Doppler (DD) domain, making the channel quasi-static and sparse. An additional benefit of OTFS in LEO-ISAC is that the round-trip sensing parameters (i.e., delay and Doppler parameters) are double of the delay and Doppler in the DD domain. Hence, these sensing parameters can be utilized to estimate the communication channels and reduce the communication overhead. The sensing and communication functions can also be jointly optimized by designing precoder or waveform in DD domain, or one function can be prioritized while the other assists.

% In addition, the delay and Doppler in the DD domain is the half of the round-trip sensing parameters (i.e., delay and Doppler parameters) in the reflected radar echo, thus the sensing parameters can be utilized to estimate the communication channels due to its quasi-static feature in the DD domain.

\textbf{Energy-efficient Transceiver Design:}
The LEO satellites are normally power limited, thus their power needs to be used with high efficiency. One possible solution is to design hybrid beamforming to reduce the total number of RF chains, which in turn reduces power consumption. Since the power consumption of electronic components (e.g., analog-to-digital converter (ADC) and phase shifter) in the RF chain is proportional to their bit budget, another possible direction is to adopt low-resolution electronic component to save the power.

\textbf{Echo-miss:}
Typically, communication frames are much longer than the radar echo round-trip times (RTTs). 
% For example, in the 5G NR specifications, a standard radio frame has $10$ms duration. For a LEO satellite located at orbits with altitude $1000$km, its echo RTT is of the order of $1-10$ms - orders of magnitude smaller than the radio frame duration. 
Thus, for a mono-static setup, the strong self interference (SI) from the transmitter can drown out the radar echo. To resolve this issue, SI cancellation techniques for LEO-ISAC system can be explored. Additionally, adjacent LEO satellites can form a bistatic radar to prevent SI.

\textbf{Multi-static LEO-ISAC System:}
Integration of multiple LEO satellites in the constellation network can function as a multi-static radar. Consequently, the multi-static LEO-ISAC system can jointly sense the targets with improved detection and tracking capabilities, increased survivability, and greater flexibility. However, the data exchange (e.g., sharing of the radar reference waveform) and coordination among LEOs need to be further investigated. One possible solution is utilizing ISL to provide the capability of data exchange among LEOs.

{
\textbf{Security:}
Due to the spectrum sharing mechanism as
well as the broadcasting nature of satellite communications,
the ISAC system is susceptible to  unique security challenges.
Since the radar target has a high reception SINR on the embedded confidential signal, it greatly amplifies the vulnerability of information eavesdropping by the target.
% it can significantly increase the susceptibility of information eavesdropping by the target.
Possible solutions include cryptographic techniques, e.g., advanced encryption algorithms, and 
physical layer (PHY) security techniques, e.g., 
secure beamforming, artificial noise approaches jamming and
cooperative security.
}

{
\textbf{Complexity and Compatibility:}
% Implementing an LEO satellite ISAC system involves specific practical considerations.
% The system can be expensive due to satellite launch, sensing and communication payloads, operation, maintenance, and upgrades over time.
In recent years, advancements in satellite miniaturization, reusable rockets, and developments in the space industry have reduced the associated costs due to satellite launch, operation, and upgrades over time.
However, the system can be inherently complex due to the management of orbital dynamics, seamless handover, precise timing synchronization, and coordination of the system.
% Satellite communication systems typically employ specific protocols and frequency bands. 
The compatibility of ISAC
with existing LEO satellite communication systems is still in its infancy stage and requires compatibility of specific protocols,  frequency bands, existing ground station infrastructure, and LEO satellite operators, etc.
More research is required to bridge the gap between theory and implementation.
}

\vspace{-0.3cm} 
\section{Conclusion}
LEO-ISAC is envisioned as a promising technique for future wireless networks due to its lower latency, density, coverage and ease of integration characteristics. This paper provides a tutorial that explores the potential of applying ISAC in LEO satellite systems for various use cases. We first analyze the current status of LEO satellite systems, and present the motivation for ISAC-enabled LEO satellite systems. To facilitate LEO-ISAC systems, we then provide potential techniques that are opportunistic and optimized ISAC techniques. RSMA is highlighted as an optimized ISAC technique for LEO-ISAC systems due to its effectiveness in simultaneously managing interference and enabling better communication-sensing trade-off performance. The development of LEO-ISAC is still in its infancy and there are many open issues to investigate before fully leveraging its advantages. Finally, the unsolved open challenges and future directions are summarized. We hope that our discussion will stimulate interest and further investigation for LEO-ISAC in future wireless networks.
%
% The aim of this activity is to investigate this phenomenon further.
\vspace{-0.2cm} 
\bibliographystyle{IEEEtran}
\bibliography{refss}

\end{document}